\documentclass[12pt]{iopart}
\usepackage[T1]{fontenc}
\usepackage[utf8]{inputenc}
\usepackage{lmodern}
\usepackage[english]{babel}
\usepackage[autostyle]{csquotes}

\usepackage{float}
\usepackage{graphics,graphicx}
\usepackage{subfigure}
\usepackage{iopams}

\expandafter\let\csname equation*\endcsname\relax
\expandafter\let\csname endequation*\endcsname\relax

\usepackage{mathtools,leftidx} 

\usepackage{xcolor}

\usepackage{booktabs}
\usepackage{multirow}
\usepackage{stmaryrd}
\usepackage{cite}

\newcommand{\He}{$\ensuremath{\prescript{3}{}{\mathrm{He}}}$}
\newcommand{\B}{$\ensuremath{\prescript{10}{}{\mathrm{B}}}$}
\newcommand{\Li}{$\ensuremath{\prescript{6}{}{\mathrm{Li}}}$}

\newcommand{\Be}{$\ensuremath{\prescript{9}{}{\mathrm{Be}}}$}

\usepackage[Euler]{upgreek}
\usepackage{textcomp}

\definecolor{aaltoOrange}{RGB}{255,121,0}%
\definecolor{aaltoBlue}{RGB}{0,101,189}%

\begin{document}

\title{Boron doping in gallium oxide from first principles}
\author{Jouko Lehtom\"aki$^1$, Jingrui Li$^1$ and Patrick Rinke$^1$}
\address{$^1$ Department of Applied Physics, Aalto University, 00076 AALTO, Finland}

\begin{abstract}
We study the feasibility of boron doping in gallium oxide ($\text{Ga}_2\text{O}_3$) for neutron detection.  $\text{Ga}_2\text{O}_3$ is a wide band gap, radiation-hard material with potential for neutron detection, if it can be doped with a neutron active element. We investigate the boron-10 isotope as possible neutron active dopant. Intrinsic and boron induced defects in $\text{Ga}_2\text{O}_3$ are studied with semi-local and hybrid density-functional-theory calculations. We find that it is possible to introduce boron into gallium sites at moderate concentrations. High concentrations of boron, however, compete with the boron-oxide formation.
\end{abstract}
\maketitle

\section{Introduction}

Gallium oxide (Ga$_2$O$_3$) is a wide gap semiconductor (band gap $E_{\text{g}}\sim$ 4.9~eV \cite{doi:10.1063/1.3521255}) with potential applications in ultraviolet optoelectronic devices, power electronics and laser lithography \cite{doi:10.1002/pssa.201330197, doi:10.1063/1.1330559, doi:10.1063/1.359055, doi:10.1063/1.5006941, doi:10.1063/1.4977766, doi:10.1063/1.5034444}.  In this work, we are exploring further applications of Ga$_2$O$_3$ for neutron detectors. There is a growing need for neutron detectors with low-power requirements, compact size and reasonable resolution for, e.g., non-invasive neutron imaging of organic materials, like human tissue or wood \cite{NeutronImaging}, safeguarding and non-proliferation of nuclear material \cite{Mukhopadhyay/etal_review:2014},  safety in the nuclear industry \cite{Caruso_review:2010}, space science \cite{Mastro01012017} and autonomous radiation probes for hazardous environments \cite{doi:10.1002/aelm.201600501}. 

Most current neutron detectors use helium-3 gas (\He), a non-radioactive isotope of helium, because of its extreme sensitivity in detecting neutron radiation \cite{Mukhopadhyay/etal_review:2014,Caruso_review:2010,milbrath_peurrung_bliss_weber_2008}. However, innovation is greatly needed, since current neutron detectors are expensive, bulky and not radiation-hard, precisely because of their use of \He. The world's \He\ supply is extremely scarce and depleting rapidly. Moreover, the large size of \He\-based detectors limits their portability and spatial resolution. Since \He\ detectors are not radiation-hard, they cannot be used in harsh environments like outer space, or fusion or nuclear reactors.  For these reasons, semiconductor detectors have recently received increasing attention \cite{Mukhopadhyay/etal_review:2014,Caruso_review:2010,milbrath_peurrung_bliss_weber_2008,melton2011development, doi:10.1063/1.4868176, doi:10.1063/1.4929913, 2012PhDT38H, Mandal2017}. However, the materials requirements for optimal energy, time and spatial resolution, detection efficiency, robustness and radiation hardness are daunting challenges \cite{milbrath_peurrung_bliss_weber_2008}, and there is currently no satisfying material choice nor commercially available semiconductor detectors. For this reason, we are here exploring Ga$_2$O$_3$ as potential neutron detector material.

Solid state neutron detectors use \textit{neutron active} elements, which convert neutrons to electronic excitation via a nuclear reaction. The ability of neutron active elements to capture neutrons is measured by the neutron cross section. The boron isotope \B\ has the largest neutron cross-section at 3840 barns, which is comparable to helium (\He) and larger than other candidates like lithium (\Li) and beryllium (\Be). 
Boron-based neutron detectors have recently been demonstrated experimentally \cite{doi:10.1063/1.4868176,2012PhDT38H, Mandal2017}, but are far from commercialisation. Wide band-gap materials have also been investigated in solid state neutron detectors, most notably gallium nitride (GaN) \cite{melton2011development, doi:10.1063/1.4929913}. Here, we consider beta gallium oxide ($\upbeta\text{-Ga}_2\text{O}_3$) as a potential material for neutron detection, because $\upbeta\text{-Ga}_2\text{O}_3$ is a radiation-hard wide band-gap material, and gallium has similar chemical characteristics as boron which makes boron implantation on gallium sites favorable.

The electronic structure of $\upbeta\text{-Ga}_2\text{O}_3$ and the behavior of defects in the material have attracted considerable interest and have been studied previously with density function theory (DFT) \cite{doi:10.1063/1.3521255,He2006, PhysRevB.87.235206, doi:10.1063/1.5019938, Tadjer01012019, doi:10.1063/1.5020134, doi:10.1063/1.5140742}. Defects have been investigated as a source of the observed intrinsic n-type conductivity and for the possibility of p-type doping of $\upbeta\text{-Ga}_2\text{O}_3$ for opto-electronic applications. Boron-related defects have not been previously studied in $\upbeta\text{-Ga}_2\text{O}_3$.

In this work we investigated the possibility of boron doping with DFT. With the supercell approach, we calculated formation energies for simple point defects and complexes in $\upbeta\text{-Ga}_2\text{O}_3$ in the diffuse doping limit. We studied both intrinsic defects and boron defects to assess the feasibility of introducing boron into $\upbeta\text{-Ga}_2\text{O}_3$. Our work provides insight into the limits of boron doping and the potential of $\upbeta\text{-Ga}_2\text{O}_3$ for neutron detection.

The article is structured as follows. Section~\ref{sec:CD} reviews briefly the atomic structure of $\upbeta\text{-Ga}_2\text{O}_3$ and outlines the computational details. In Section~\ref{sec:results} we discuss the results of DFT calculations with a particular focus on boron doping in $\upbeta\text{-Ga}_2\text{O}_3$. Section~\ref{sec:conclusion} concludes with a summary.

\section{Computational Details}
\label{sec:CD}

\begin{figure}
    \centering
    \includegraphics[width=\textwidth]{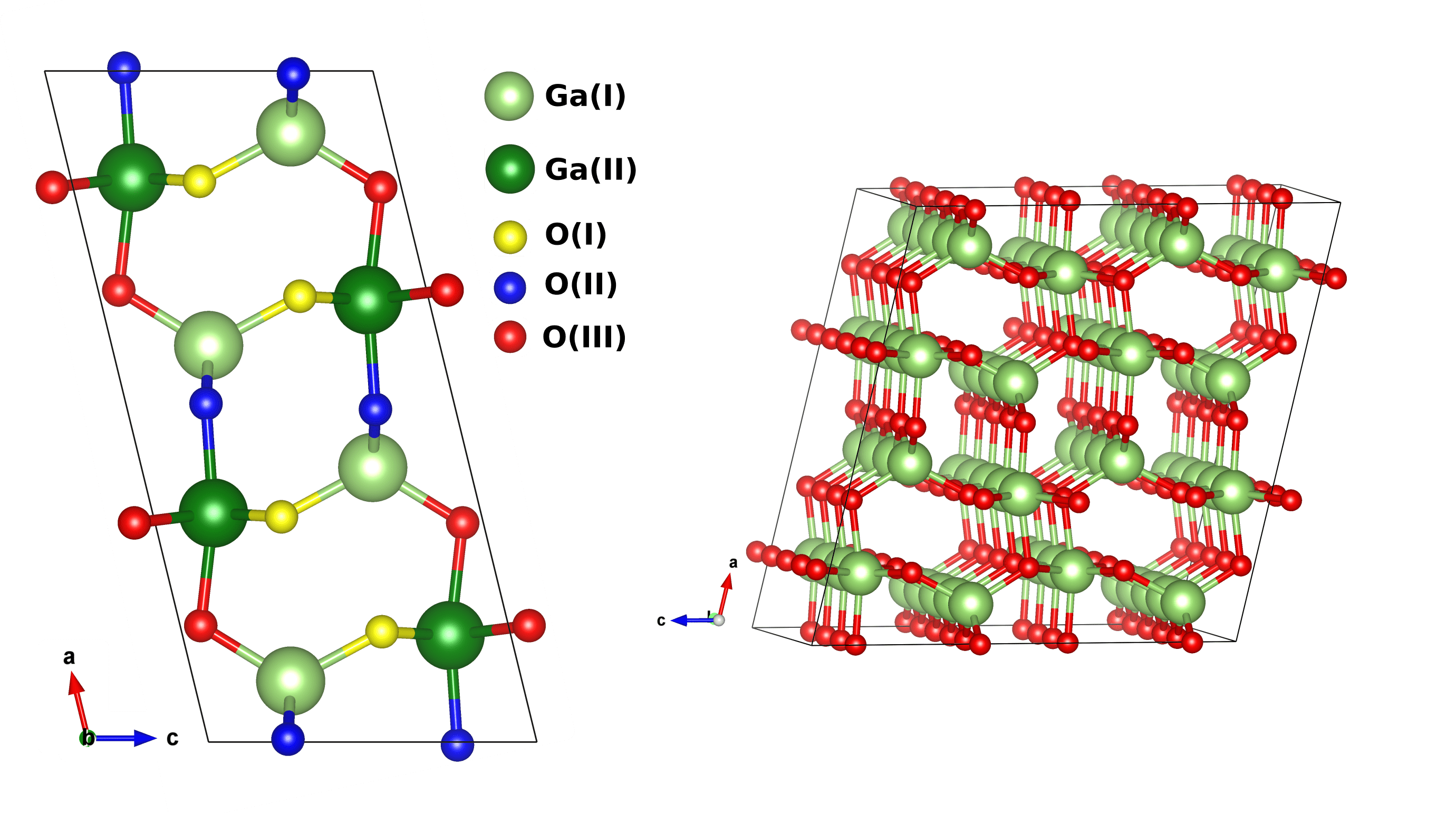}
    \caption{Left: Conventional monoclinic unit cell of $\upbeta\text{-Ga}_2\text{O}_3$. All nonequivalent gallium and oxygen sites are color-coded. Right: Supercell of 160-atoms constructed from the conventional unit cell.
    } 
    \label{fig:unit_cell}
\end{figure}

$\upbeta\text{-Ga}_2\text{O}_3$ has a monoclinic crystal structure with space group C2/$m$.
The unit cell contains two nonequivalent gallium sites and three nonequivalent oxygen sites. The monoclinic cell with 4 $\text{Ga}_2\text{O}_3$ units (i.e., 20 atoms) is shown in Fig.~\ref{fig:unit_cell}. The five different sites are labeled as Ga(I), Ga(II), O(I), O(II) and O(III). The gallium sites Ga(I) and Ga(II) are tetrahedrally and octahedrally coordinated by O ions, respectively. The O(III) site is four-fold coordinated, while both O(I) and O(II) are three-fold coordinated. An O(I) site has two Ga(II) and one Ga(I) as neighbors, while an O(II) has two Ga(I) and one Ga(II) neighboring sites.

All defect calculations were carried out with the supercell approach \cite{RevModPhys.86.253} in this work. Point defects were introduced in a 160-atom supercell model of pristine $\upbeta\text{-Ga}_2\text{O}_3$, i.e., 32 $\text{Ga}_2\text{O}_3$ units. 
Following Ref.~\cite{RevModPhys.86.253}, we calculated the defect formation energy according to
\begin{align}
\label{eq:Eform}
    E_{\text{f}}(\text{X}^q) = & E(\text{X}^q) - E(0) + E_{\text{corr}} + q(\epsilon_\text{VBM} + \epsilon_{\text{F}}) - \sum_i \Delta n_i \mu_i ,
\end{align}
where $E(\text{X}^q)$ is the DFT total energy of the supercell containing a defect in charge state $q$, and $E(0)$ the total energy of the defect-free crystal. $\mu_i$ is the chemical potential of the $i$th species whose number varies by $\Delta n_i$ when defects are formed. $\Delta n_i$ is negative for the removal of atoms (e.g., vacancies) and positive for the addition of atoms (e.g., interstitials).  $\epsilon_{\text{F}}$ is the Fermi energy of Ga$_2$O$_3$, defined with respect to the valance band maximum ($\epsilon_{\text{VBM}}$). The $q(\epsilon_{\text{VBM}}+\epsilon_{\text{F}})$ term therefore accounts for the energy change upon  removal or addition of electrons when charge defects are formed. 

To remove spurious electrostatic interactions between supercells with charged defects, we included the Freysoldt-Neugebauer-\mbox{Van~de~Walle} (FNV) correction term $E_\text{corr}$  \cite{PhysRevLett.102.016402}. In the FNV scheme, we used a spatially averaged dielectric constant of $\epsilon_0 \sim 10$ \cite{doi:10.1063/1.359055,Varley2010} which includes ionic and electronic screening \cite{doi:10.1063/1.5054826}. There has been some debate, if the electronic dielectric constant $\epsilon_\infty$ should be used instead for small supercells \cite{Deak2017}. However, we observed that $\epsilon_0$ is the correct choice by extrapolating supercells to the infinite supercell limit (see Appendix~\ref{sec:appendix-FNV}). Our findings are in agreement with those of  Ingebrigtsen et al. \cite{doi:10.1063/1.5054826}.

The chemical potentials for species $i$ can be written as
\begin{align*}
    \mu_{i} = \mu^0_i + \Delta \mu_i,
\end{align*}
where $\Delta \mu_i \leq 0$ and $\mu^0_i$ is acquired from $T = 0$ K DFT calculation of the appropriate phase, e.g. gas phase of the O$_2$ molecule for O, and solid metal Ga with space group Cmce for gallium. We incorporated the external environment through the temperature and partial pressure dependence of the chemical potentials of the gas-phase species, i.e., here only oxygen
\begin{align}
\label{eq:t_p_dependence}
    \Delta \mu_\text{O}(T,p) = \frac{1}{2}\left\{\left[ H_0 + \Delta H(T)\right] - T\left[ S_0 + \Delta S(T) \right]\right\} + \frac{1}{2}k_{\text{B}} T\ln\left(\frac{p}{p_0}\right).
\end{align}
Here $H_0$ and $S_0$ are enthalpy and entropy at zero temperature, respectively. All values were referenced to 1 atm pressure and obtained from thermodynamic tables~\cite{https://doi.org/10.18434/t42s31}.

We estimated the boron doping concentration $c$ in various conditions with the Arrhenius relation \cite{PhysRevB.65.035406}
\begin{align}
\label{eq:arrhenius}
    c(\text{X}^q) = N_\text{site}N_\text{config}\exp\left(-G_\text{f}(\text{X}^q)/k_{\text{B}}T\right),
\end{align}
where $\text{X}^q$ is the configuration of a boron dopant, $N_\text{site}$ the number of dopant sites per unit volume and $N_\text{config}$ their configurational degeneracy factor. The Gibbs free energy is approximated as
\begin{align}
\label{eq:Gibbs_free_energy}
    G_{\text{f}}(\text{X}^q) \approx E(\text{X}^q) - E(0) + E_{\text{corr}} + q(\epsilon_\text{VBM} + \epsilon_{\text{F}}) - \sum_i \Delta n_i \mu_i\left( T, p\right),
\end{align}
where $\mu_\text{Ga}(T,p) \approx \mu_\text{Ga}$, but for oxygen we use $\mu_\text{O}(T,p) = \mu^0_\text{O} + \Delta\mu_\text{O}(T,p)$ from eq.~\eqref{eq:t_p_dependence}. With this approximation, we took into account only the pressure- and temperature-dependence of the oxygen chemical potential and discarding other entropy contributions from the bulk phases. Note that this is very simplistic approximation for the Gibbs free energy as it is almost the same as the formation energy (eq.~\eqref{eq:Eform}) but still useful \cite{doi:10.1063/1.5019938}. With this approximation, the only difference between the Gibbs free energy and the zero temperature formation energy is that the gas-phase chemical potentials have a temperature- and pressure-dependence via the ideal gas relation.

All DFT calculations in this work were performed with the all-electron numeric-atom-centered orbital code \textsc{fhi-aims} \cite{FHI_aims,HavuV09,Ren_2012,LEVCHENKO201560}. We used the semi-local Perdew-Burke-Ernzerhof (PBE) functional \cite{Perdew96} and the Heyd-Scuseria-Ernzerhof hybrid functional (HSE06) \cite{doi:10.1063/1.1564060}  to calculate the atomic and electronic structure of $\upbeta\text{-Ga}_2\text{O}_3$ and defects therein. PBE calculations were employed as reference to previous work and to test the supercell dependence for charge corrections. For the final defect geometries, we always used the HSE06 functional to avoid spurious delocalization effects in PBE, as observed for, e.g., the oxygen vacancies in TiO$_2$ \cite{Janotti/etal:2010}. We set the fraction of Hartree-Fock exchange in HSE06 to $35\%$, a value which has been previously used for Ga$_2$O$_3$ \cite{Varley2010}. This yields a band gap of $4.95$~eV for tight settings in FHI-aims and $4.76$~eV for light settings (see below for these two settings), thus providing an acceptable compromise between accuracy and computational cost. Scalar relativistic effects were included by means of the zero-order regular approximation (ZORA) \cite{vanLenthe93}.

Considering the computational cost of HSE06 calculations, we carried out most of our calculations with the cheaper ``light'' basis sets (which usually provide sufficiently converged energy differences) and used results with ``tight'' basis sets (which can better provide converged absolute energies) as reference. For light settings, we used the tier-1 basis set for oxygen and gallium, but exclude the $f$ function for gallium. For tight settings,  we use tier-2 for oxygen and the full tier-1 basis for gallium. Adding tier 2 for gallium did not improve the result for PBE. The tier-1 basis set for gallium is therefore enough to achieve convergence. A $\Gamma$-centered $2\!\times\!8\!\times\!4$ $k$-point mesh was used for the 20-atom monoclinic unit-cell calculations, while for larger supercells (160-atom) we used a $\Gamma$-centered $2\!\times\!2\!\times\!2$ $k$-point mesh. In pursuit of open materials science \cite{MatDatPerspective}, we made the results of all relevant calculations available on the Novel Materials Discovery (NOMAD) repository \cite{nomad_repo}.

\section{Results}
\label{sec:results}

\subsection{Bulk Ga$_2$O$_3$ and chemical potentials}

The optimized geometry of bulk $\upbeta\text{-Ga}_2\text{O}_3$ is presented in Table~\ref{tab:geometry_beta} for the HSE06 and PBE functionals. Band gaps and formation enthalpies have been included for completeness. The PBE functional overestimates the lattice constants compared to experiment. Conversely, the HSE06 functional reproduces the experimental geometry well and our results are consistent with those previously reported in the literature \cite{He2006, Varley2010, PhysRevB.87.235206, Deak2017, RevModPhys.86.253}. 

\begin{figure}
    \centering
    \includegraphics[width=0.7\textwidth]{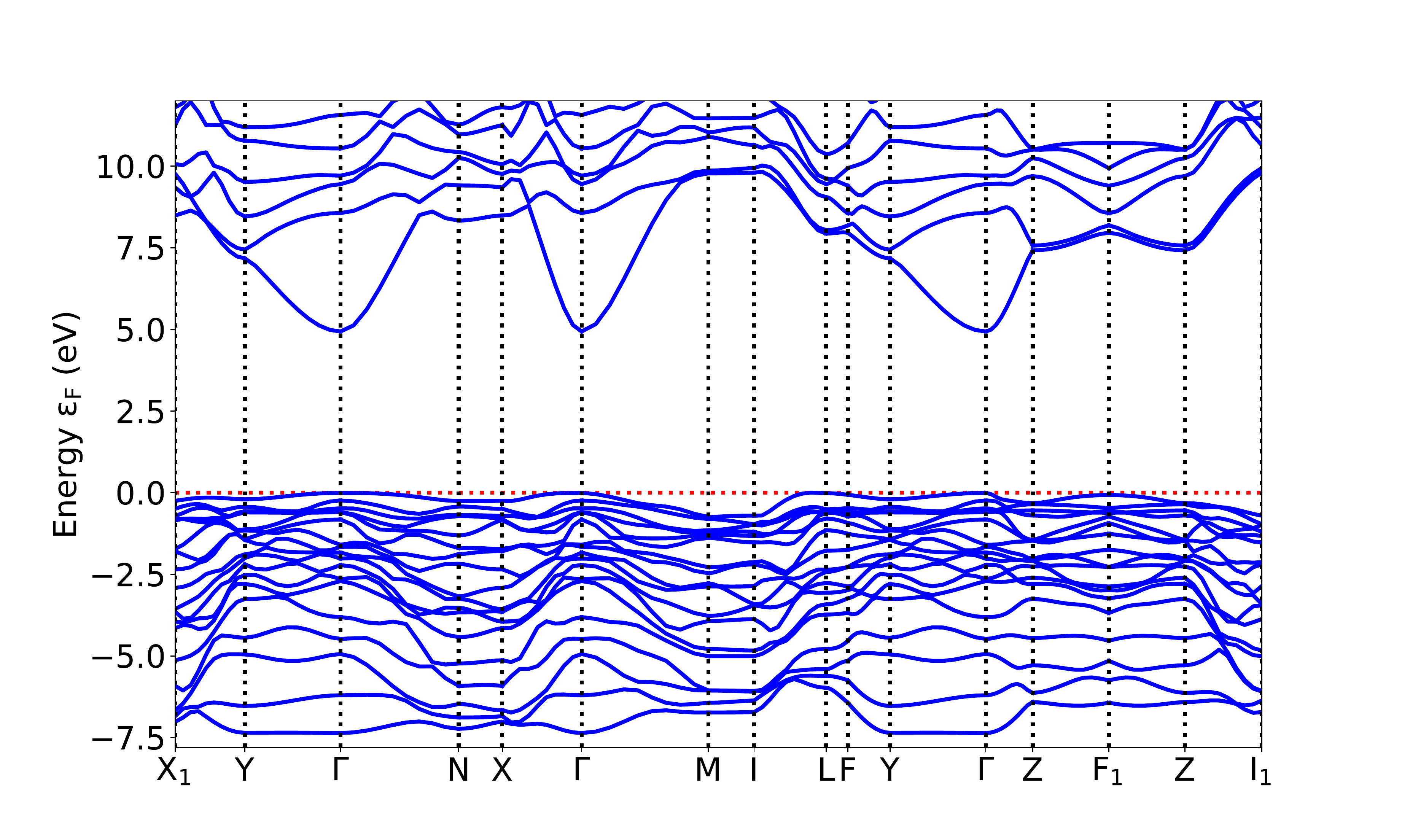}
    \caption{HSE06 band structure of $\upbeta$-$\text{Ga}_2\text{O}_3$ along the  path defined in Ref.~\cite{doi:10.1002/pssb.201451551}.}
    \label{fig:band_structure}
\end{figure}

The HSE06 band structure of $\upbeta\text{-Ga}_2\text{O}_3$ is shown in Fig.~\ref{fig:band_structure}. The band gap of 4.92~eV is indirect between a point in the I-L line for the VBM and the $\Gamma$-point for the conduction band minimum (CBM). The direct gap at the $\Gamma$-point is slightly larger (4.95~eV). The fact that indirect transitions are weak makes $\upbeta\text{-Ga}_2\text{O}_3$ effectively a direct band-gap material.

\begin{table}[!ht]
\centering
\caption{Lattice parameters ($a$, $b$, $c$ and $\beta$) of bulk $\upbeta\text{-Ga}_2\text{O}_3$, as well as the band gap ($E_{\text{g}}$) and formation energy ($H_{\text{f}}$) calculated with different DFT functionals. $H_{\text{f}}$ is given in eV per $\text{Ga}_2\text{O}_3$ unit. Also listed are experimental (Exp.) results for the lattice parameters  \cite{doi:10.1063/1.1731237} and band gap \cite{doi:10.1063/1.3521255} as reference. 
}
\label{tab:geometry_beta}
\begin{tabular}{@{\hspace{0em}}r@{\hspace{0.5em}}crc@{\hspace{0.5em}}crc@{\hspace{0.5em}}crl@{\hspace{0em}}}
\midrule\midrule
& \multicolumn{3}{c}{PBE} & \multicolumn{3}{c}{HSE06} & \multicolumn{3}{@{\hspace{0.5em}}c@{\hspace{0em}}}{Exp.} \\ \midrule
$a~[\text{\AA}]$ & & $12.46$ & & & $12.23$ & & & $12.23$ & \cite{doi:10.1063/1.1731237}\\
$b~[\text{\AA}]$ & &  $3.08$ & & &  $3.05$ & & &  $3.04$ & \cite{doi:10.1063/1.1731237}\\
$c~[\text{\AA}]$ & &  $5.88$ & & &  $5.81$ & & &  $5.80$ & \cite{doi:10.1063/1.1731237}\\
$\beta~[\text{\textdegree}]$ & & $103.7$ & & & $103.7$ & & & $103.7$ & \cite{doi:10.1063/1.1731237} \\ \hline
$E_{\text{g}}~[\text{eV}]$ & &  1.95 & & &  4.95 & & &  ~4.9 & \cite{doi:10.1063/1.3521255} \\
$H_{\text{f}}~[\text{eV}]$ & & -10.6 & & & -10.1 & & & -11.3 & \cite{doi:10.1063/1.1731237} \\ \midrule\midrule
\end{tabular}
\end{table}

We reference the gallium chemical potential $\mu^0_\text{Ga}$ to gallium metal and the oxygen chemical potential $\mu^0_\text{O}$ to the oxygen molecule $\text{O}_2$ (see \ref{sec:app_chempot} for details). The chemical potentials need to be in equilibrium (i.e, $2\mu_\text{Ga} + 3\mu_\text{O} = E(\text{Ga}_2\text{O}_3)$), which defines the Ga-rich ($\Delta\mu_\text{Ga} = 0$) and O-rich ($\Delta\mu_\text{O} = 0$) limits. 
An important constraint on the boron chemical potential is the formation of boron oxide $\text{B}_2\text{O}_3$. The upper bound of the boron chemical potential is therefore $2\mu_\text{B} + 3\mu_\text{O} \leq E(\text{B}_2\text{O}_3)$.  We use solid boron as the boron chemical potential $\mu_{B}^0$.

\subsection{Intrinsic defects}
\label{sec:intrinsic_defects}
We first investigate intrinsic point defects. We do this not only to validate our calculations against previous studies, but also to study the competition between intrinsic defects and boron defects. Here we present only vacancy sites while in \ref{sec:app_intrinisc_defects} we provide calculations for other relevant intrinsic defects \cite{Deak2017, doi:10.1063/1.5054826}. 

The most important transition states of vacancy defects are listed in Table~\ref{tab:vacancy_transition_levels}. The charge transition levels of the oxygen vacancies  $\epsilon(+2/0)$ are located deep below the CBM. Different coordinations yield slightly different transition states with the four-fold O(III) site being closest to the CBM. For n-type conditions (Fermi energy close to the CBM), the oxygen vacancies are therefore neutral while they would behave as donors for p-type conditions (Fermi energy close to the VBM). Conversely, gallium vacancies act as deep acceptors for most of the Fermi energy range. Here the  $\epsilon(-2/-3)$ transition state for the lower coordinated Ga(I) is closer to the CBM than the octahedral Ga(II) state. We note in passing, that the Ga(I) vacancy in the -2 charge state requires a hybrid functional treatment. In the PBE functional the extra electrons do not localize, resulting in a formation energy that is too low.

\begin{table}[!ht]
\centering
\caption{Transition levels of vacancy defects. All energies (in eV) are given with respect to the conduction band minimum (CBM). The transition level is the energy at which two defect charge states, $q$ and $q'$, are in equilibrium. Reference~\cite{doi:10.1063/1.5054826} uses 32\% fraction of exact exchange in HSE06 while in Ref.~\cite{Deak2017} 26\% exact exchange is used with no range separation.} 
\label{tab:vacancy_transition_levels}
\begin{tabular}{@{\hspace{0em}}l@{\hspace{1em}}c@{\hspace{1em}}c@{\hspace{1em}}c@{\hspace{1.0em}}c@{\hspace{0em}}}
\midrule\midrule
Vacancy & \multirow{2}{*}{$q/q'$} & \multicolumn{3}{c}{Transition level}  \\
\multicolumn{1}{c}{site}    &        & This work & Ref.~\cite{doi:10.1063/1.5054826} & Ref.~\cite{Deak2017}\\ \midrule
Ga(I)  & (-2/-3)    & -1.65 & -1.76 & -1.64 \\
Ga(I)  & (-1/-2)    & -2.21 & -2.32 &   -\\
Ga(II) & (-2/-3)    & -2.06 & -2.17 & -2.12\\
Ga(II) & (-1/-2)    & -2.39 & -2.50 &  -\\
O(I)   & (+2/0)     & -1.38 & -1.50 & -1.71\\
O(II)  & (+2/0)     & -2.11 & -2.23 & -2.29\\
O(III) & (+2/0)     & -1.24 & -1.36 & -1.56\\ \midrule\midrule
\end{tabular}
\end{table}

Our results  agree qualitatively and quantitatively with the existing literature for simple vacancy defects.  Our transition levels are consistently lower than those reported in Ref.~\cite{doi:10.1063/1.5054826}, which is most likely due to the different amount of exact exchange in the HSE06 functional (32 \% in Ref.~\cite{doi:10.1063/1.5054826} and 35 \% in this work) and therefore a different bulk band gap of Ga$_2$O$_3$. On the experimental side, efforts are ongoing to identify point defects in Ga$_2$O$_3$ \cite{Deak2017, doi:10.1063/1.5054826}. However, thus far, no clear assignments have been possible.

\subsection{Boron defects}
\label{sec:boron_defects}
Next we turn to boron point defects. We did initial calculations for neutral defects with the PBE functional, which are shown in \ref{sec:app_boron}.  PBE and HSE06 give the same formation energy ordering for neutral defects. We therefore scanned a variety of neutral defects with PBE. A clear picture emerges:  4-fold coordinated boron defects are the lowest in energy. We then picked three substitutional defects on Ga-sites with one or two borons and further investigated them with HSE06.

\begin{figure*}
    \centering
    \includegraphics[width=\textwidth]{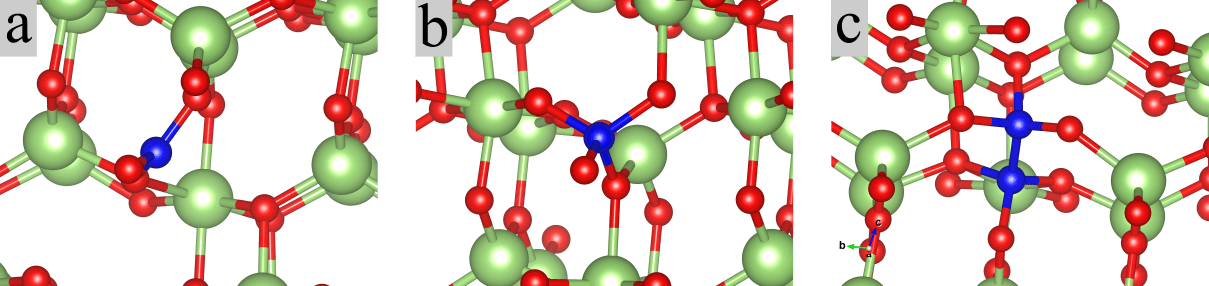}
    \caption{Structure of the boron defect sites in $\text{Ga}_2\text{O}_3$ supercell: (a) Boron on Ga(II)-site with three-fold coordination, (b) boron on Ga(I)-site with four-fold coordination c) two 4-fold coordinated boron atoms on the Ga(II) site. Ga, O and B atoms are colored in light green, red and blue, respectively.}
    \label{fig:defect_structures}
\end{figure*}

\begin{figure*}
    \centering
    \includegraphics[width = \textwidth]{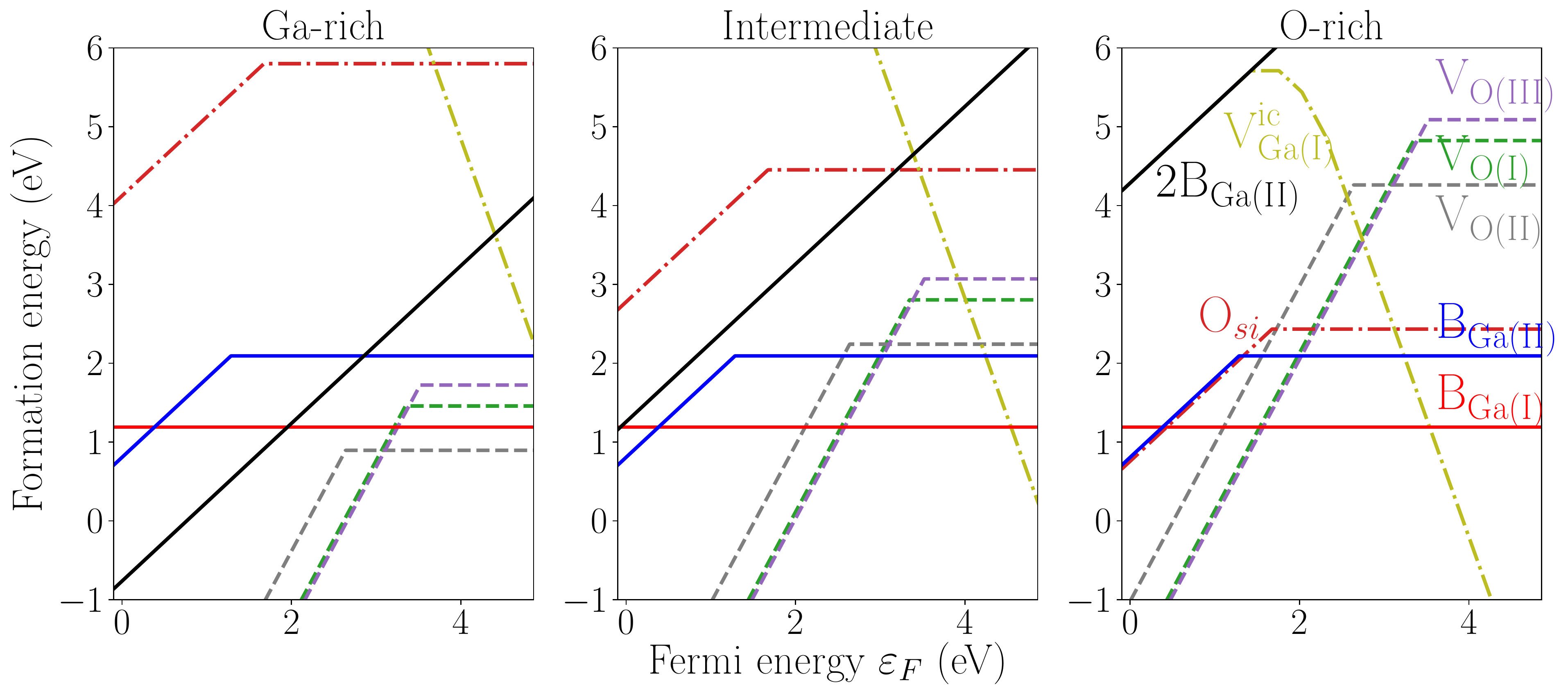}
    \caption{Defect formation energies $E_f$ for multiple intrinsic defects and the most important boron defects. The chemical potential of Gallium is $\mu_\text{Ga} = \frac{1}{5} H_{\text{f}}(\text{Ga}_2\text{O}_3)$ for the intermediate case. The boron chemical environment is set to $\Delta\mu_\text{B} = -1.28 + \Delta\mu_\text{Ga}$ which gives the lowest possible formation energies for boron related defects while preventing formation of $\text{B}_2\text{O}_3$. See text for more details.}
    \label{fig:defect_diagram}
\end{figure*}

The boron defect geometries are shown in Fig.~\ref{fig:defect_structures} and the corresponding formation energies in Fig.~\ref{fig:defect_diagram} for three different chemical environments (O-rich, Ga-rich and intermediate conditions $\mu_\text{Ga} = \mu_\text{O} = \frac{1}{5} H_{\text{f}}(\text{Ga}_2\text{O}_3)$). Boron preferably incorporates into the tetrahedrally coordinated Ga(I) site. The neutral $\text{B}_\text{Ga(I)}$ substitutional defect is very stable and does not introduce charge states into the band gap. Boron on the Ga(II) site, $\text{B}_\text{Ga(II)}$, is not able to maintain the 6-fold coordination of the substituted gallium due to its much smaller ionic size. This leads to a larger relaxation of the surrounding atoms such that $\text{B}_\text{Ga(II)}$ becomes 3-fold coordinated and introduces a dangling bond on one of the neighboring oxygen atoms. In this site, boron can therefore act as donor with a $\varepsilon(+1/0)$ transition state  at $1.29$~eV above the VBM. 

Another interesting boron defect is the two-boron complex on the Ga(II) site ($2\text{B}_\text{Ga(II)}$) shown in Fig.~\ref{fig:defect_structures}. Each boron is 4-fold coordinated, which makes the formation energy competitive to the other two boron defects we discussed. Similar two boron structures were constructed on the Ga(I) and interstitial sites but they were not 4-fold coordinated thus resulting in considerably higher formation energies.

Next, we address the range of boron chemical potential, in which boron defects form preferentially.
By combining the equilibrium condition of $\upbeta\text{-Ga}_2\text{O}_3$ and the restriction of $\text{B}_2\text{O}_3$ formation on the boron and oxygen chemical potentials, we arrive at $\Delta\mu_\text{B} - \Delta\mu_\text{Ga} \leq \frac{1}{2}\left[H_{\text{f}}(\text{B}_2\text{O}_3) - H_{\text{f}}(\text{Ga}_2\text{O}_3)\right] = -1.28$ eV, where $H_\text{f}$ is the heat of formation. The implication is that to prevent the formation of $\text{B}_2\text{O}_3$, the chemical potential of boron must always be lower than that of gallium $\mu_\text{B} \leq \mu_\text{Ga}$. Thus the most boron rich environment is $\Delta \mu_\text{B} = -1.28\text{~eV} + \Delta\mu_\text{Ga}$.

In Fig.~\ref{fig:defect_diagram} we show intrinsic defects and boron defects  in different chemical environments, for which the boron chemical potential obeys  $\Delta \mu_\text{B} = -1.28\text{~eV} + \Delta\mu_\text{Ga}$. 
Clearly the incorporation of neutral borons on gallium sites, especially Ga(I), is the most preferable way of doping. Boron complexes with multiple boron atoms are not favored, since the penalty term of not forming  $\text{B}_2\text{O}_3$ suppresses them. Furthermore, neutral boron defects are preferable as we are not interested in making electronically active defects, but incorporating boron as a neutron active material.

\subsection{Boron doping}
We now perform a semi-quantitative analysis of boron doping based on the boron defects on gallium sites. Our main goal is to ascertain, if we can introduce significant concentrations of boron for neutron detection. We are aiming for a boron concentration of $10^{22}$~cm$^{-3}$. This number is estimated based on experiments performed for GaN \cite{doi:10.1063/1.4868176}, which demonstrated neutron detection in boron-doped GaN for a boron density of $5.12\times10^{22}$~cm$^{-3}$. The density of \B\ is $10^{22}$ $\times$ cm$^{-3}$ considering a \B\ abundance of $\sim 20 \%$. In addition, GaN itself is neutron-active due to the presence of N, which implies that we would probably require more B implantation than our estimate.

For the substitutional boron defects, $N_\text{config}$ in the Arrhenius relation in eq.~\eqref{eq:arrhenius} is equal to 1 and the site density $N_\text{site}$ for both gallium sites is $1.92 \times 10^{22} \text{cm}^{-3}$.  Inserting the formation energies of the boron defects shown in Fig.~\ref{fig:defect_diagram} into the Arrhenius relation reveals that the concentration ratio  $\text{B}_\text{Ga(II)}/\text{B}_\text{Ga(I)} = \exp\left(\left[E\left(\text{B}_\text{Ga(I)}\right) - E\left(\text{B}_\text{Ga(II)}\right) \right]/k_{\text{B}} T\right)$  is between $2.6\times 10^{-8}$ and $6.8\times 10^{-3}$ for temperatures between 600 K and 2100 K, which is the relevant range for doping and $\text{Ga}_2\text{O}_3$ crystal growth. We therefore only consider $\text{B}_\text{Ga(I)}$ in the following.  Similarly, $2\text{B}_\text{Ga(II)}$ is also excluded from further consideration as it has a considerably higher formation energy than $\text{B}_\text{Ga(II)}$ in all chemical environments where the formation of B$_2$O$_3$ is unfavorable.

First we investigate the boron concentrations as a function of temperature in a chemical environment optimal for boron implantation. The formation energy of $\text{B}_\text{Ga(I)}$ depends on the chemical environment through  the difference in the gallium and boron chemical potential $\mu_\text{Ga} - \mu_\text{B}$. This is further constrained by the formation of the competing B$_2$O$_3$ phase, which results in the inequality $\Delta\mu_\text{Ga} - \Delta\mu_\text{B} \geq \frac{1}{2}\left[H_{\text{f}}(\text{Ga}_2\text{O}_3) - H_{\text{f}}(\text{B}_2\text{O}_3)\right] = 1.28$ eV that guarantees that the formation of B$_2$O$_3$ is unfavorable. 

\begin{figure*}
    \centering
    \includegraphics[width=0.7\textwidth]{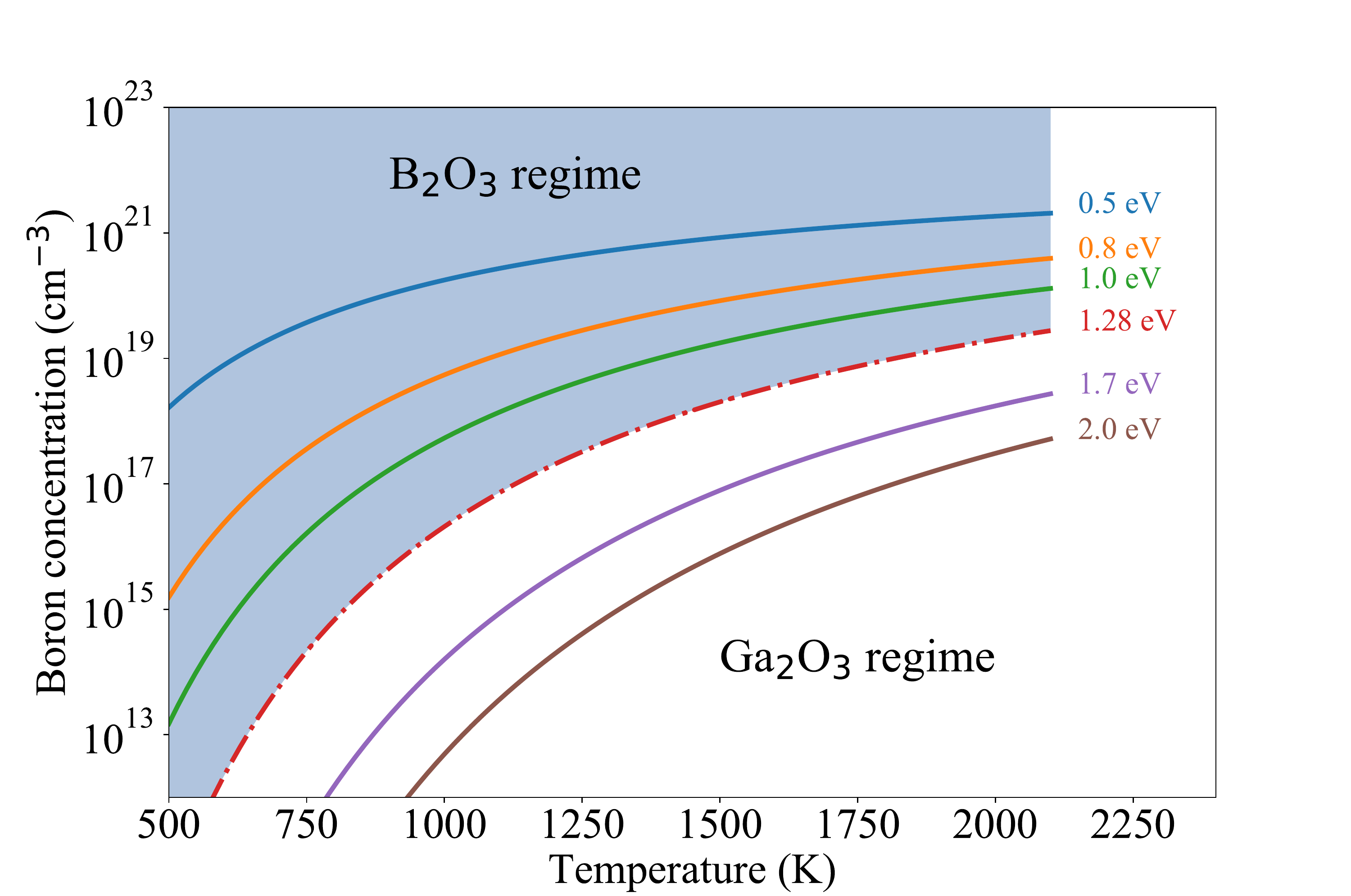}
    \caption{Concentration of boron defect $\text{B}_\text{Ga(I)}$ as a function of growth temperature where different lines have different chemical environments described in terms of difference $\Delta \mu_\text{Ga} - \Delta \mu_\text{B}$. Concentrations are calculated with the Arrhenius relation (eq.~\eqref{eq:arrhenius}). Boron oxide is a limiting factor $\Delta \mu_\text{Ga} - \Delta \mu_\text{B} \geq 1.28$ eV marked with a dashed line.
    } 
    \label{fig:c_vs_T}
\end{figure*}

In Fig.~\ref{fig:c_vs_T}, we plot the boron concentrations for different chemical environments as a function of temperature for growth temperatures from 600 K up to 2100 K. Higher temperatures favor boron incorporation and the boron concentration increases with growth temperature. Furthermore, boron rich conditions (i.e. small values of $\Delta \mu_\text{Ga} - \Delta \mu_\text{B}$) are more conducive to boron incorporation than gallium rich (high values). Unfortunately, the divider line of $\Delta \mu_\text{Ga} - \Delta \mu_\text{B}=1.28 eV$ implies that in reality the B dopability might be quite low. Even at the highest crystal growth temperatures we are limited to a boron concentration of $2.0 \times 10^{19} \text{cm}^{-3}$ ($\sim 0.2$ \% of the total Ga(I) sites) and are thus quite far away from our goal of $10^{22}$~cm$^{-3}$. Growth methods that extend into the B$_2$O$_3$ regime, but suppress the formation of boron oxide, would be beneficial. 

\begin{figure*}
    \centering
    \includegraphics[width=\textwidth]{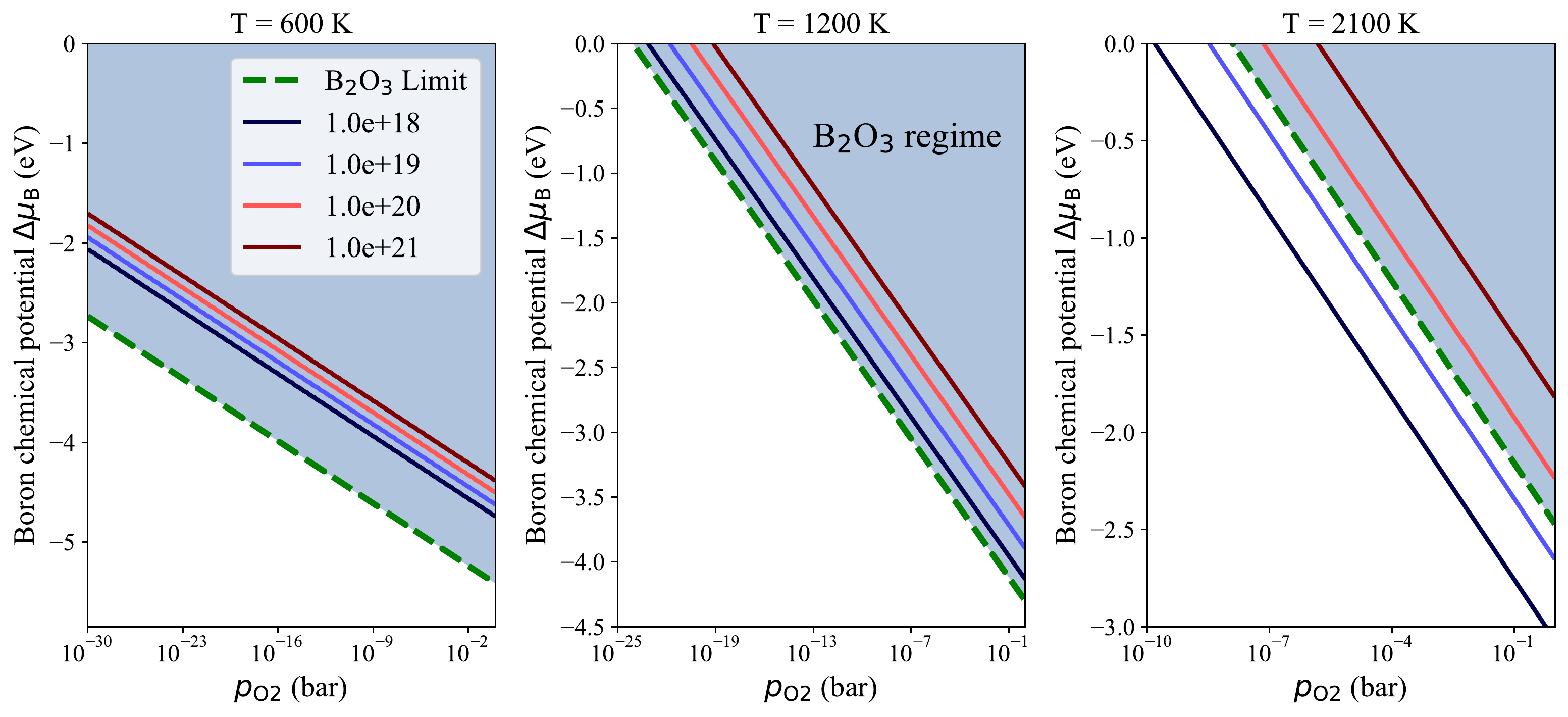}
    \caption{Contours of $\text{B}_\text{Ga(I)}$ concentration as a function of oxygen partial pressure given by eq.~\eqref{eq:t_p_dependence}. Limiting boron oxide is marked as a dashed green line and the area favorable for B$_2$O$_3$ formation is marked with light grey. }
    \label{fig:contour_pO2_muB}
\end{figure*}

Finally, we make a connection between the boron chemical potential and the oxygen environment. In Fig.~\ref{fig:contour_pO2_muB}, we plot the boron chemical potential as a function of the oxygen partial pressure. We do not convert $\Delta \mu_\text{B}$ into a partial pressure, since boron may not be supplied in pure gaseous form during growth. Figure~\ref{fig:contour_pO2_muB} shows that, if we are targeting a certain boron concentration  (straight lines), $\Delta \mu_\text{B}$ has to reduce with increasing oxygen partial pressure. The relation arises from the fact that the gallium chemical potential is tied to the oxygen chemical potential via equilibrium conditions. The boron concentration depends on Gibbs free energy \eqref{eq:Gibbs_free_energy} via Arrhenius relation \eqref{eq:arrhenius} where the chemical potentials are $\Delta\mu_\text{Ga} - \Delta\mu_\text{B}$ which can be then transformed into expression $\frac{1}{2}H_\text{f}(\text{Ga}_2\text{O}_3) -\frac{3}{2}\Delta\mu_\text{O} - \Delta\mu_\text{B}$ via equilibrium condition of gallium and oxygen chemical potential. Higher partial pressures imply higher oxygen chemical potential, and in order to keep the boron concentration constant, the boron chemical potential has to be lowered.

As a side note, there is a distinct possibility that low $\text{O}_2$ pressures are not accessible due to formation of gallium suboxide (Ga$_2$O) which makes $\upbeta\text{-Ga}_2\text{O}_3$ unstable \cite{doi:10.1063/1.4942002, doi:10.1063/1.5019938}. A possible formation of gallium suboxide would depend on the growth method and  we do not explore this phenomenon further in this context.

In Fig.~\ref{fig:contour_pO2_muB} we also marked the B$_2$O$_3$ growth regime. It is apparent that meaningful boron concentrations fall into this B$_2$O$_3$ regime at lower growth temperatures. Only at $1200$ K and above we can obtain reasonable concentrations near the B$_2$O$_3$ limit. Acquiring even boron concentrations of $1.0\times 10^{20} \text{cm}^{-3}$ ($\sim 1\%$ of Ga(I) sites) would require going above the B$_2$O$_3$ limit even for high temperatures. To stress the limitation, we calculated the required partial pressures with boron gas B$_2$ as the boron reference. For a temperature of 1200 K and oxygen partial pressures $p(\text{O}_2)$ above $10^{-30}$ bar, the partial pressure of B$_2$ would have to be below $10^{-40}$ bar: the boron environment would have to be extremely poor even in oxygen poor conditions, which are also limited due to stability of gallium oxide. Such low amounts of boron or oxygen would also limit the growth/doping rate.

From these results, it is apparent, that it is challenging to introduce high concentrations of boron into $\upbeta\text{-Ga}_2\text{O}_3$ without formation of B$_2$O$_3$. For neutron detectors it is possible to enhance the neutron activity by constructing thicker layers of the material to obtain a higher number of neutron active atoms, but here we do not explore technical device details. Compared to previous experimental results (5.12 $\times10^{22}$ cm$^{-3}$), achievable boron concentrations appear to be quite moderate in Ga$_2$O$_3$ according to our calculations.

\section{Conclusion}
\label{sec:conclusion}
We have investigated boron related point defects in  $\upbeta\text{-Ga}_2\text{O}_3$ with DFT for a possible use of the material in solid-state neutron detectors. We found that boron preferably incorporates onto 4-fold coordinated gallium sites. Such boron defects are electronically neutral and do not introduce trap states in the band gap. Larger boron complexes have similar formation energies, but are unlikely due to their competition with $\text{B}_2\text{O}_3$ formation. The Ga-rich growth regime turns out to be the most conducive to boron incorporation.

Boron can be introduced as a substitutional defect to gallium sites in meaningful concentrations, but the concentrations are still modest compared to previous boron-based neutron active materials, mostly due to the limitations imposed by B$_2$O$_3$. The limitation might likely inhibit introducing boron also to other oxide materials such as In$_2$O$_3$. The situation would be improved, if growth methods could be extended into the B$_2$O$_3$ stability region. 

\section*{Acknowledgment}

We thank F. Tuomisto, V. Havu, S. Kokott and D. Golze for fruitful discussions. The generous allocation of computing resources by the CSC-IT Center for Science (via Project No.~ay6311) and the Aalto Science-IT project are gratefully acknowledged. This work was supported by the Academy of Finland through its Centres of Excellence Programme under project number 284621, as well as its Key Project Funding scheme under project number 305632.

\appendix

\section{Chemical potentials}
\label{sec:app_chempot}
For completeness, Table~\ref{tab:chemical_potentials} lists the DFT-calculated energies of several relavant systems which were used for calculating the chemical potentials. For gallium, we used Ga metal in the orthorhombic structure with 8 atoms per unit cell as reference. The reference for oxygen is the O$_2$ molecule. Boron is referenced to its $\alpha$ phase with a rhombohedral crystal structure with 12 atoms in a unit cell. For boron oxide ($\text{B}_2\text{O}_3$), we took the $\alpha$-phase with 15 atoms per unit cell \cite{Gurr:a07482}. The calculations for Ga, B and $\text{B}_2\text{O}_3$ were carried out using a $8\times8\times8$, $2\times2\times2$ and $4\times4\times4$ $\Gamma$-centered $k$-point mesh. 
\begin{table}[!ht]
\centering
\caption{Reference systems used in the calculations of the chemical potentials. For each system the energy is given per formula unit except for gallium and oxygen where it is given per atom. 
}
\label{tab:chemical_potentials}
\begin{tabular}{l@{\hspace{1.0em}}r@{\hspace{1.0em}}|lr}
\midrule\midrule
System                  & Energy (eV)   & System & Energy (eV)         \\ \midrule
$\text{Ga}$             & -53183.059  & $\text{B}$             &  -676.609 \\
$\text{O}$              & -2046.547   & $\text{B}_2\text{O}_3$ & -7505.521 \\
$\text{Ga}_2\text{O}_3$ & -112515.856 &  \\ \midrule
\end{tabular}
\end{table}

\section{Intrinsic defects}
\label{sec:app_intrinisc_defects}
\begin{figure*}[!ht]
    \centering
    \includegraphics[width=\textwidth]{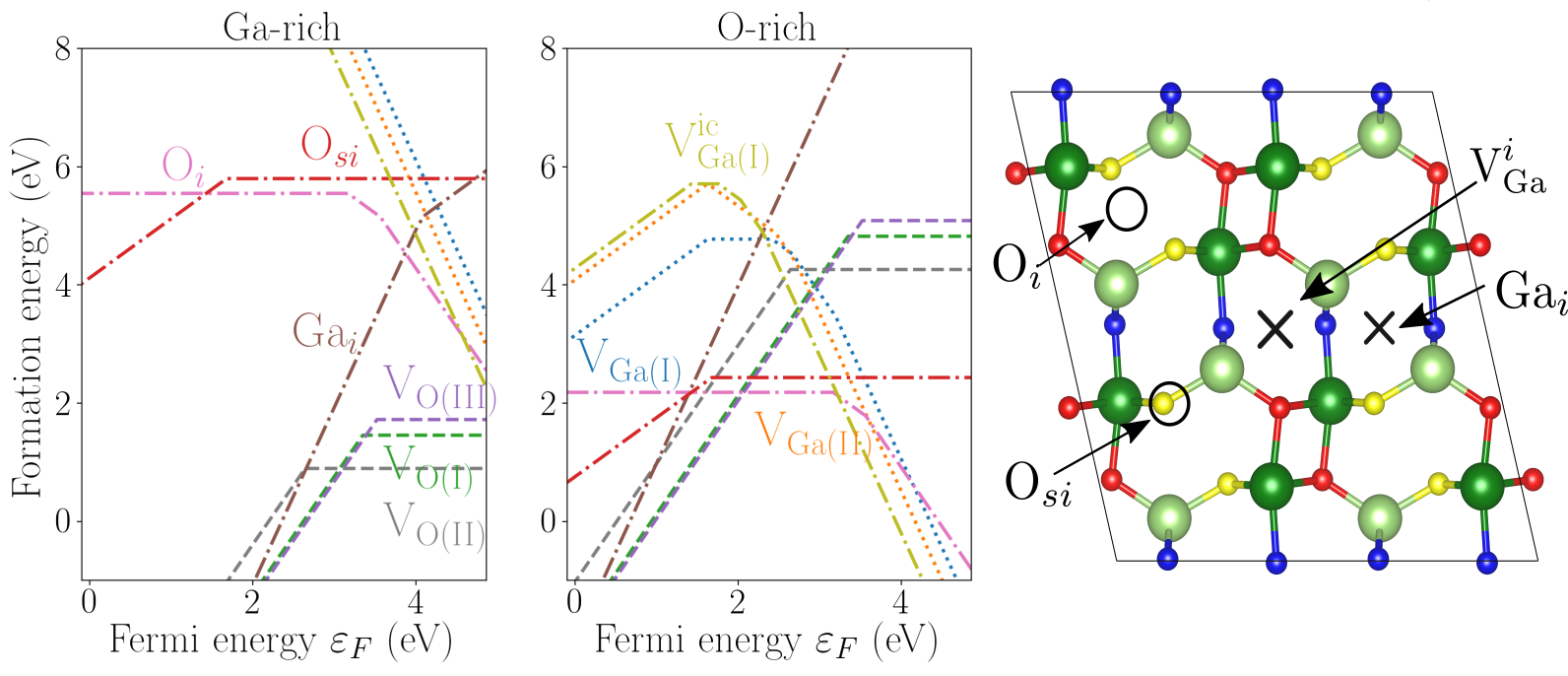}
    \caption{Formation energies for intrinsic vacancy and interstitial defects in $\upbeta\text{-Ga}_2\text{O}_3$ for Ga-rich (left) and O-rich (middle) conditions as a function of the Fermi energy. The interstitial locations are shown on the right.}
    \label{fig:intrinsic_defects}
\end{figure*}

The interstitial defects in $\text{Ga}_2\text{O}_3$ are more complex than the single vacancies (see  Fig.~\ref{fig:intrinsic_defects}). We studied two oxygen interstitials, a split interstitial ($\text{O}_{\text{si}}$) on the O(I) site and a three-fold coordinated interstitial ($\text{O}_{\text{i}}$). For gallium interstitials, we considered two different configurations. In the $\text{V}_\text{Ga}^{\text{i}}$ interstitial one gallium is removed from the Ga(I)-site and the second Ga(I) moves to an interstitial position with octahedral coordination. In the second configuration ($\text{Ga}_{\text{i}}$) we add one gallium atom with octahedral coordination into an interstitial position such that two nearby Ga(I) gallium atoms are pushed away from the interstitial gallium. The chosen transition levels are listed in Table~\ref{tab:interstitial_transition_levels}. From these defects only $\text{Ga}_{\text{i}}$ is donor-like near CBM while the gallium interstitial $\text{V}_\text{Ga}^{\text{i}}$ is similar to simpler gallium vacancies and acts as a deep acceptor for most of the Fermi energy range. The interstitial configurations are shown in \ref{sec:interstitial_configurations}. $\text{V}_\text{Ga}^\text{i}$ can be considered as defect complex of a gallium vacancy and an interstitial but we have labeled it as an interstitial because the defect is more complex than the straightforward vacancy defects in Table~\ref{tab:vacancy_transition_levels}. Our results  agree qualitatively and quantitatively with the existing literature for both intrinsic vacancy and interstitial defects, see Ref.~\cite{doi:10.1063/1.5054826}. 

\begin{table}[!ht]
\centering
\caption{Transition levels of interstitial defects. All energies (in eV) are given with respect to the CBM.}
\label{tab:interstitial_transition_levels}
\begin{tabular}{l@{\hspace{0.5em}}c@{\hspace{0.5em}}c@{\hspace{0.5em}}c}
\midrule\midrule
\multirow{2}{*}{Defect} & \multirow{2}{*}{$q/q'$} & \multicolumn{2}{c}{Transition level} \\
  & & This work & \cite{doi:10.1063/1.5054826} \\ \midrule
$\text{O}_{si}$        & (+1/0)  & -3.08     & -3.26  \\
$\text{O}_i$           & (-1/-2) & -1.20     & -1.23  \\
$\text{V}_\text{Ga}^i$ & (-2/-3) & -2.46     & -2.55  \\
$\text{V}_\text{Ga}^i$ & (-1/-2) & -2.73     & -2.82  \\
$\text{V}_\text{Ga}^i$ & (0/-1)  & -3.00     & -3.29  \\
$\text{Ga}_i$          & (+3/+1) & -0.69     & -0.60 
\\ \midrule\midrule
\end{tabular}
\end{table}

\section{Boron defects with the PBE functional}
\label{sec:app_boron}
In Table~\ref{tab:neutral_pbe_defects} we tabulate neutral defects calculated with the PBE and HSE06 functional. The formation energies are given for the Ga-rich ($\mu_\text{Ga} = 0$~eV) and boron rich ($\mu_\text{B} = -1.17$~eV) limit. The gallium and oxygen vacancies are listed for reference to demonstrate that they have the same energetic ordering as neutral vacancies with the HSE06 functional. 

Boron defects $\text{B}_\text{Ga(I)}$ and $\text{B}_\text{Ga(II)}$ are substitutional defects on Ga-sites. More complex substitutional defects are $(2\text{B})_\text{Ga(II)}$, $(2\text{B})_\text{Ga(I)}$ and $(3\text{B})-(2\text{Ga(II)})$, in  which two or three boron atoms replace Ga atoms. The $\text{B}_{\text{i}}$ interstitial has a lower formation energy than the $(2\text{B})-\text{Ga}_{\text{i}}$ interstitial, in which a gallium atom moves to an interstitial site and the vacant Ga-site is filled with two substitutional borons.

\begin{table}[!ht]
\centering
\caption{The formation energies (eV) of neutral Boron defects and vacancies computed with PBE and HSE06 functional. See text for details.}
\label{tab:neutral_pbe_defects}
\begin{tabular}{l@{\hspace{1.0em}}crc@{\hspace{1.0em}}crc}
\midrule\midrule
Defect                         & \multicolumn{3}{c@{\hspace{1.0em}}}{$E_f$ (PBE)} & \multicolumn{3}{c}{$E_f$ (HSE06)} \\ \midrule
$\text{V}_\text{Ga(I)}$        && 9.483       &&& 9.825         \\
$\text{V}_\text{Ga(II)}$       && 9.597       &&& 10.755        \\
$\text{V}_\text{O(I)}$         && 1.085       &&& 4.824         \\
$\text{V}_\text{O(II)}$        && 0.607       &&& 4.262         \\
$\text{V}_\text{O(III)}$       && 1.322       &&& 5.089         \\
$\text{B}_\text{Ga(I)}$        && -0.196      &&& 1.240         \\
$\text{B}_\text{Ga(II)}$       && 0.628       &&& 2.143         \\
$(2\text{B})_\text{Ga(II)}$    && 2.440       &&& 3.392         \\
$\text{B}_i$                   && 2.841       &&& \multicolumn{1}{c}{-}             \\
$(2\text{B})_\text{Ga(I)}$     && 3.285       &&& \multicolumn{1}{c}{-}             \\
$(3\text{B})-(2\text{Ga(II)})$ && 4.344       &&& \multicolumn{1}{c}{-}             \\
$(2\text{B})-\text{Ga}_i$      && 4.965       &&& \multicolumn{1}{c}{-}             \\ \midrule\midrule
\end{tabular}
\end{table}

\section{Electrostatic corrections}
\label{sec:appendix-FNV}
\begin{figure}
    \centering
    \includegraphics[width=0.48\textwidth]{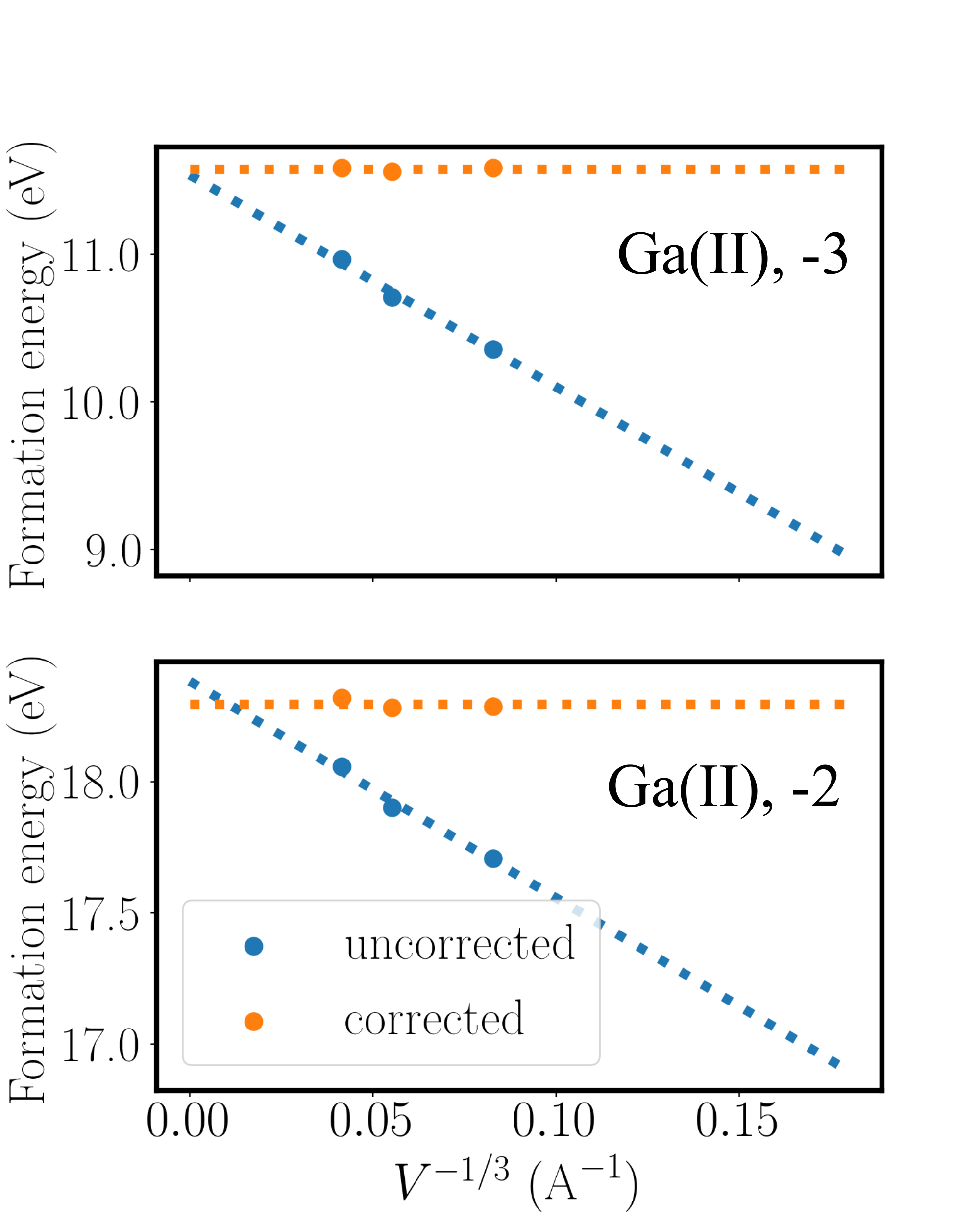}
    \caption{Defect formation energies for $\text{V}_\text{Ga(II)}$ in the $-3$ charge state (upper) and the $-2$ charge state (lower) as a function of the inverse super cell volume. Defect formation energies (symbols) are calculated for supercells of different sizes with and without the FNV correction. Lines are linear fits to the data.}
    \label{fig:charge_corrections}
\end{figure}

We verified the FNV corrections for the Ga(II) vacancy in two charge states by an explicit supercell convergence with the PBE functional. The results are shown in Fig.~\ref{fig:charge_corrections}. The structures are multiples of the unit cell, which have been relaxed after the removal of one gallium in the Ga(II)-site. For The FNV correction we use a  dielectric constant $\varepsilon_0$ of 10. Applying the FNV correction results in horizontal lines with formation energies that are independent of the supercell size.

\section{Interstitial defects in $\upbeta\text{-Ga}_2\text{O}_3$}
\label{sec:interstitial_configurations}
\begin{figure*}
    \centering
    \includegraphics[width=\textwidth]{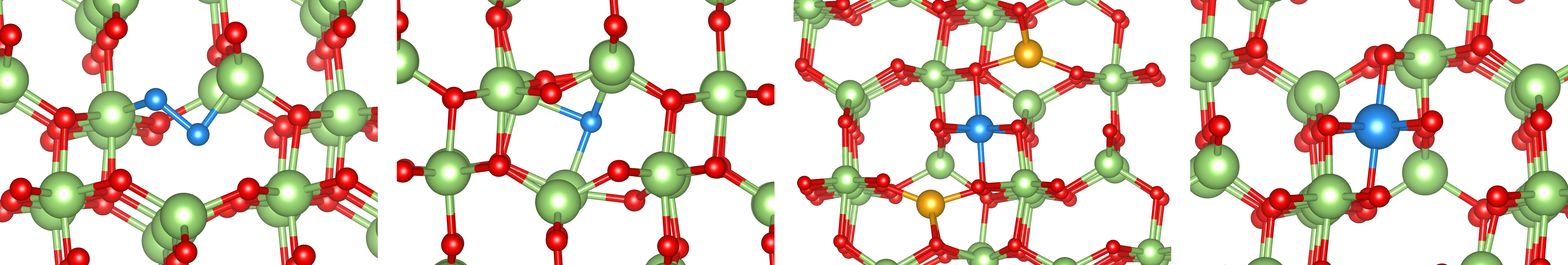}
    \caption{Atomic configurations of four different interstitials. The interstitials are marked with light blue while important deformations near the interstitial are marked with orange. From left to right: Oxygen split interstitial $\text{O}_{si}$, oxygen interstitial $\text{O}_i$ and gallium interstitial $\text{Ga}_i$ with two gallium atoms in orange which have moved from Ga(I)-sites. Finally, on the right gallium interstitial $\text{V}_\text{Ga}^i$ which is surrounded by two Ga(II)-sites and two empty Ga(I)-sites.}
    \label{fig:interstitials}
\end{figure*}

In Fig.~\ref{fig:interstitials} we show the atomic configurations for the interstitial defects. The structure of vacancies is straightforward and therefore not shown for brevity. 

\newpage

\section*{References}
\bibliographystyle{iopart-num}
\bibliography{bibliography}

\end{document}